\documentclass
[prl,notitlepage,superscriptaddress,longbibliography,reprint]{revtex4-1}
\usepackage[T1]{fontenc}
\usepackage{times}
\usepackage{xcolor}
\usepackage{graphicx}
\usepackage{bm}
\usepackage{dsfont}
\usepackage{mathrsfs}
\usepackage{eucal}
\usepackage{amsmath}
\usepackage{amsfonts}
\usepackage{amssymb}
\usepackage[colorlinks,linkcolor=blue,citecolor=blue]{hyperref}%

\renewcommand{\vec}[1]{{\ensuremath{\bm{\mathrm{#1}}}}}

\begin{document}

\title{Quantum thermodynamics of complex ferrimagnets}
\author{Joseph Barker}
\affiliation{Institute for Materials Research, Tohoku University, Sendai
980-8577, Japan}

\author{Gerrit~E.W.~Bauer}
\affiliation{Institute for Materials Research \& AIMR \& CSRN, Tohoku University, Sendai
980-8577, Japan}
\affiliation{Zernike Institute for Advanced Materials, University of
Groningen, 9747 AG Groningen, The Netherlands}

\begin{abstract}
High-quality magnets such as yttrium iron garnet (YIG) are electrically insulating and very complex. By implementing a quantum thermostat into atomistic spin dynamics we compute YIG's key thermodynamic properties, viz. the magnon power spectrum and specific heat, for a large temperature range. The results differ (sometimes spectacularly) from simple models and classical statistics, but agree with available experimental data.
\end{abstract}
\maketitle

\paragraph{Introduction}

The spin dynamics of electrically insulating magnets often has high quality
because the dissipation channel by conduction electron scattering is absent.
With few exceptions, they are complex ferrimagnets. Yttrium iron garnet (YIG)
with 80 atoms in the unit cell rules with a record low Gilbert damping of long
wavelength spin wave excitations or magnons, even at room
temperature~\cite{Cherepanov1993,Wu2013e}. The implied exceptionally low
disorder and weak coupling with phonons remains a mystery, however. Recently,
magnon heat and spin transport were measured in YIG thin films in a non-local
spin injection and detection configuration with Pt contacts by means of the
spin Hall effect~\cite{Cornelissen2015} and modelled by spin diffusion
\cite{Cornelissen2016c}. Key parameters of this model are linked to the
thermodynamics of the magnetic order, such as the magnon heat capacity, which
is difficult to measure because it is orders of magnitude smaller than the
phonon heat capacity---at 10~K the magnon and phonon heat capacities are
$C_{m}~\approx0.009$ J~kg$^{-1}$~K$^{-1}$ and $C_{p}\approx0.270$ J~kg$^{-1}%
$~K$^{-1}$~\cite{Boona2014}. They can be separated by magnetic freeze-out of
the magnon contribution at temperatures up to a few
Kelvin~\cite{Douglass1963a, Boona2014}. The magnon heat capacity at higher
temperatures has been estimated by extrapolating models that agree with
experimental low-temperature results \cite{Rezende2015,Cornelissen2016c}. YIG
is often treated as a single-mode ferromagnet with quadratic $\omega
\propto\mathcal{D}k^{2}$ (or isotropic cosine function) dispersion, thereby
ignoring higher frequency acoustic and optical modes and temperature
dependence of the exchange stiffness $\mathcal{D}$. Furthermore, magnon-magnon
interactions are also commonly neglected or treated in a mean field
approximation. Statistical approaches also have issues, such as the use of
classical (Johnson-Nyquist) thermal noise at low temperatures
\cite{Oitmaa2009}.

In this Letter we introduce a numerical method that avoids all of these
shortcomings. It allows us to carry out material-dependent thermodynamic
calculations that are quantitatively accurate with a small number of
parameters that can be determined independently. The crucial ingredient is a
thermostat for Planck quantum (rather than Rayleigh--Jeans classical)
statistics in an atomistic spin dynamics framework \cite{Barker2016b}.

With the inclusion of quantum thermal statistics we find quantitative
agreement for YIG with available experiments at low temperatures. The computed
spin wave dispersion as a function of temperature agree well with results from
neutron scattering. This low temperature quantitative benchmarking imbues
trust in the technique for calculating thermodynamic functions and allows
access to quantities such as the magnon heat capacity at room temperature that
turns out to be an order of magnitude larger than previous estimates.

\paragraph{Method}

We address the thermodynamics by computing the atomistic spin dynamics in the
long (ergodic) time limit to generate canonical ensembles of spins. The
magnetic moments (`spins') in this model are treated as classical unit vectors
$\vec{S}$, an excellent approximation for the half-filled 3d-shell of the iron
cations in YIG with $\mathcal{S}=5/2$ and magnetic moment $\mu_{s}=g\mu
_{B}\mathcal{S}$, where $g\approx2$ is the electron $g$-factor and $\mu_{B}$
the Bohr magneton.

The Heisenberg Hamiltonian $\mathscr{H}=-\frac{1}{2}\sum_{ij}J_{ij}\vec{S}%
_{i}\cdot\vec{S}_{j}$ contains the (super)-exchange parameters $J_{ij}$
between spins on sites $i$ and $j$, which are determined by fits to inelastic
neutron scattering data \cite{Plant1977}. Recently, the magnon dispersions
were measured again with higher resolution~\cite{Princep2017a}, allowing an
improved parameterization of the six nearest-neighbors exchange constants,
which we adopt in the following. We add a Zeeman term $\mathscr{H}=-\sum
_{i}\mu_{s,i}\vec{H}_{\mathrm{ext}}\cdot\vec{S}_{i}$ with $\vec{H}_{\mathrm{ext}}=H_{z}=0.1~\mathrm{T}$,
to fix the quantization axis. On each lattice site `$i$' the spin dynamics
obey the Landau-Lifshitz equation of motion~\footnote{We prefer the
Landau-Lifshitz rather than the Gilbert damping, because latter affects the
frequencies (here governed exclusively by the exchange parameters) by a factor
$1/(1+\eta^{2})$.}:
\begin{equation}
\frac{\partial\vec{S}_{i}}{\partial t}=-|\gamma|\left(  \vec{S}_{i}\times
\vec{H}_{i}+\eta\vec{S}_{i}\times(\vec{S}_{i}\times\vec{H}_{i})\right)
\mathrm{,} \label{eq:llg}%
\end{equation}
where $\gamma=g\mu_{B}/\hbar$ is the gyromagnetic ratio and $\eta$ is a
damping constant. Each spin feels an effective magnetic field $\vec{H}%
_{i}=\vec{\xi}_{i}-(1/\mu_{s,i})\partial\mathscr{H}/{\partial\vec{S}_{i},}$
where $\vec{\xi}_{i}$ are stochastic processes controlled by the thermostat at
temperature $T$. $\langle\xi_{i\alpha}\rangle=0$ and the correlation function
in frequency space is governed by the fluctuation-dissipation theorem (FDT)
$\langle\xi_{i\alpha}\xi_{j\beta}\rangle_{\omega}=2\eta\delta_{ij}%
\delta_{\alpha\beta}\varphi(\omega,T)/\mu_{s,i},$ where the Kronecker
$\delta^{\prime}$s reflect the assumption that the fluctuations between
lattice sites $i,j$ and Cartesian coordinates $\alpha,\beta$ are uncorrelated.
$\varphi(\omega,T)$ describes the temperature dependence of the noise power
and is chosen such that the steady-state distribution functions obey
equilibrium thermal statistics. By not approximating the spin Hamiltonian by a
truncated Holstein-Primakoff expansion, our approach includes magnon-magnon
interactions to all orders \cite{Kittel1963}.

Atomistic spin dynamics methods generally assume the classical limit of the
FDT with frequency independent (white) noise $\varphi(\omega,T)=k_{B}T$, i.e.
\emph{all} magnons are stimulated. The energy equipartition of the coupled
system results in the Rayleigh-Jeans magnon distribution. However, this is
only valid when the thermal energy is much larger than that of the magnon mode
$k$ under consideration, i.e. when $k_{B}T\gg\hbar\omega_{k}$, while the
energies of the YIG magnon spectrum--and that of most room temperature
magnets--extend up to $\hbar\omega_{\vec{k}}/k_{B}\approx1000$%
~K~\cite{Cherepanov1993}. A classical thermostat therefore generates too many
high energy magnons, which, for example, overestimates the broadening by
magnon scattering and leads to other predictions that can be very wrong.

In the proper quantum FDT~\cite{LandauStatPhys}
\begin{equation}
\varphi(\omega,T)=\sum_{k}\frac{\hbar\omega_{k}}{\mathrm{exp}\left(
\hbar\omega_{k}/k_{B}T\right)-1},\label{eq:qfdt}%
\end{equation}
which means that equipartition is replaced by Planck statistics of the magnons
at temperature $T$. Quantum statistics in classical spin systems can partially
be mimicked through a post-process rescaling of the
temperature~\cite{Evans2015a} or by using temperature-dependent frequencies
that rely on analytic expressions for the low temperature
spectrum~\cite{Woo2015}. These approaches cannot be used to evaluate all
thermodynamic properties and are not suitable to treat complex magnets such as
YIG. We therefore adopt here the `quantum thermostat' as introduced earlier in
molecular dynamics \cite{Dammak2009,Savin2012}, i.e., a correlated noise
source that obeys the quantum FDT. This is a \textquotedblleft
colored\textquotedblright\ noise, but very different from the one used to
describe classical memory effects in the heat
bath~\cite{Atxitia2009,Ruckriegel2015}.

We implement the quantum statistics by generating correlated fluctuating
fields $\vec{\xi}_{i}\left(  t\right)  $ numerically in time that obey the FDT
in the frequency domain. Savin et al.~\cite{Savin2012} employ a set of
stochastic differential equations that produce the required distribution
function. We adjust this method for the spin dynamics problem, but can refer
the reader to Ref.~\cite{Savin2012} for the technical details. The solution
provides a dimensionless stochastic process $\Phi_{i\alpha}(t)$ with the
spectrum of Eq.~\eqref{eq:qfdt}. The dimensionful noise in the spin dynamics
reads
\begin{equation}
\xi_{i\alpha}(t)=k_{B}T\sqrt{\frac{2\eta\mu_{s,i}}{\gamma\hbar}}\Phi_{i\alpha
}(t).
\end{equation}
When we agitate the model of classical spins with these stochastic fields, the
excitations of the ground state (magnons) obey quantum statistics, quite
analogous to quantized phonons in a classical ball-spring lattice. This
approach may loosely be called a `semi-quantum' method which should work very
well for the large Fe$^{3+}$ spin in YIG with $\mathcal{S}=5/2$, but requires
more scrutiny for spin $\mathcal{S}=1/2$.

We integrate equation~(\ref{eq:llg}) using the Heun method with time step
$\Delta t = 0.1$~fs. The stochastic differential equations of the thermostat
are integrated using the fourth-order Runge-Kutta method with the same time step.

\paragraph{Magnon spectrum}

\begin{figure}[ptb]
\includegraphics[width=0.48\textwidth]{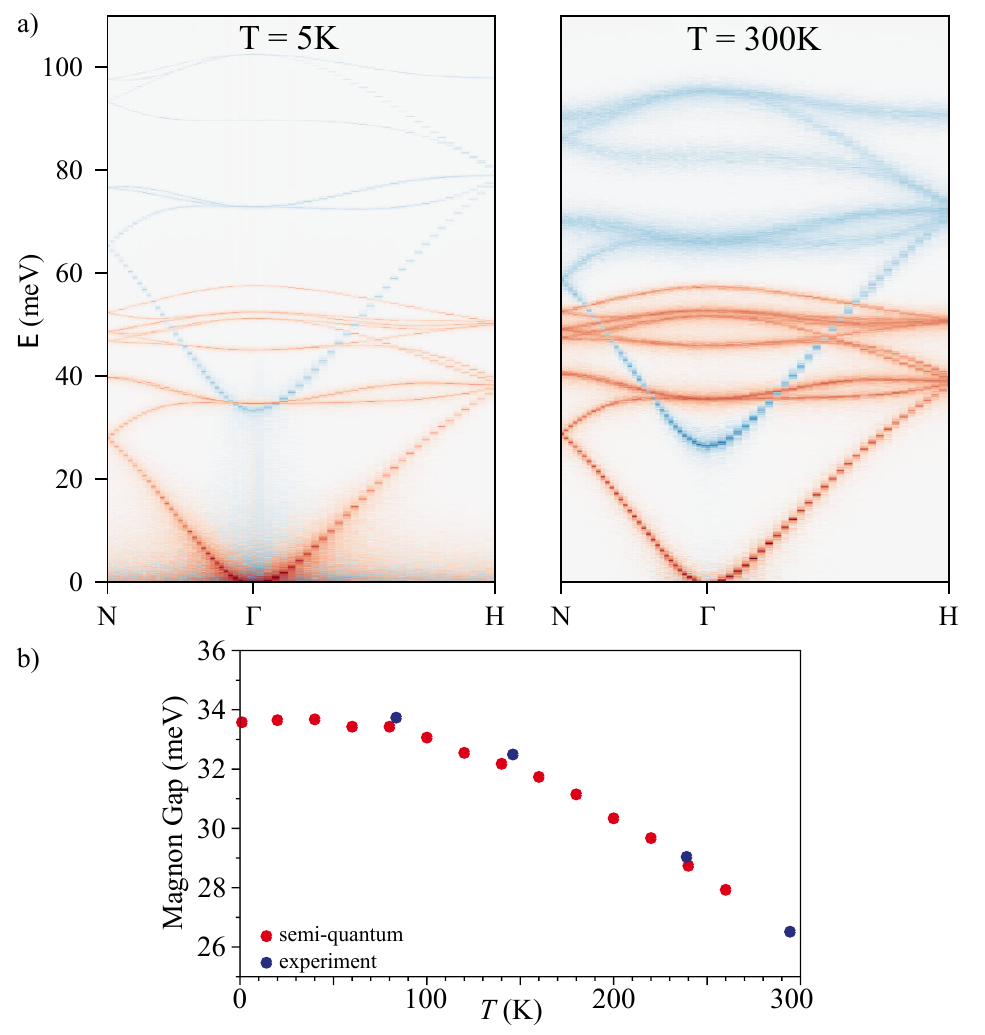}\caption{a)
YIG magnon spectrum at $T=5$~K and $T=300$~K calculated using the quantum
thermostat and the exchange parameters of Ref.~ \onlinecite{Princep2017a}. The
color intensity is adjusted on a log scale such that all modes are visible
(even for extremely low occupation) and is different for both figures. The
red/blue color shows the +/- polarization of the magnons. b) Magnon gap
between optical and acoustic modes at $\Gamma$. Experimental data are adopted
from neutron scattering experiments \cite{Plant1977}.}%
\label{fig:yig_spectrum}%
\end{figure}

We compare now the magnon spectrum computed with the quantum thermostat with
our previous work with classical statistics (and older exchange constants from
Ref.~\cite{Cherepanov1993})~\cite{Barker2016b}. Results for low ($5$~K) and
room ($300$~K) temperature are shown in Fig.~\ref{fig:yig_spectrum}a. The
classical thermostat overestimated the number of high-energy magnons and
therefore the broadening of the optical modes at higher temperatures. With
quantum statistics, the high-energy optical modes are well resolved at room
temperature and should be observable by inelastic neutron scattering with
large frequency transfer. The agreement between the calculated and measured
\cite{Plant1977} temperature dependence of the exchange gap between optical
and acoustic modes at the $\Gamma$ point, shown in Fig.~\ref{fig:yig_spectrum}%
b, is improved, especially in the low temperature regime.

\paragraph{Magnetization}

The magnetization at low temperatures $m_{z}=1-\tfrac{1}{\mathcal{S}}%
\sum_{\vec{k}\nu}\langle n_{\vec{k}\nu}\rangle_{T},$ where $\langle n_{\vec
{k}\nu}\rangle_{T}$ is the distribution of magnons with wave vector $\vec{k}$
and band index $\nu$ in the first Brillouin zone, cannot be calculated
correctly with classical statistics~\cite{Kuzmin2005} (at higher temperatures
the expression does not hold since magnon-magnon interactions are important).
This is obvious already for the single parabolic band, non-interacting magnon
gas model for which
\begin{equation}
1-m_{z}(T)=v_{\mathrm{ws}}\frac{1}{\mathcal{S}}\frac{{\Gamma\left(  \tfrac
{3}{2}\right)  }{\zeta\left(  \tfrac{3}{2}\right)  }}{2\pi^{2}}\left(
\frac{k_{B}T}{\mathcal{D}}\right)  ^{3/2}\label{eq:bloch}%
\end{equation}
where $\omega_{k}=\mathcal{D}k^{2},$ spin-wave stiffness $\mathcal{D}%
=2\mathcal{S}\mathcal{J}a^{2},$ lattice constant $a,$ $v_{\mathrm{ws}}$ volume
of the Wigner-Seitz cell, while $\Gamma(x)$ and $\zeta(x)$ are the gamma and
Riemann zeta functions. The $T^{3/2}$ dependence is known as Bloch's
law~\cite{Bloch1930}.

In the ferrimagnet YIG the total magnetization is made up by two oppositely
aligned sublattices with slightly different temperature dependent
magnetizations. At low temperatures they are rigidly locked to an antiparallel
configuration by the strong nearest neighbor exchange. At energies
$\hbar\omega_{\vec{k}} / k_{B}\lessapprox30~\mathrm{K}$ YIG's magnon dispersion is known to be
quadratic and its magnetization obeys Bloch's $T^{3/2}$
law~\cite{Srivastava1987}. The expected deviations at higher temperatures can
be assessed by our method. We calculate the magnetization at temperature $T$
as an average $\langle\cdots\rangle_{T}$ over the spin configurations at many
times over a 1~ns trajectory $\vec{m}(T)=\langle N^{-1}\sum_{i}^{N}\mu
_{s,i}\vec{S}_{i}\rangle_{T}/\langle N^{-1}\sum_{i}^{N}\mu_{s,i}\vec{S}%
_{i}\rangle_{T=0}$, where $N=655,360$ is the total number of spins in the simulation.

\begin{figure}[ptb]
\includegraphics[width=0.48\textwidth]{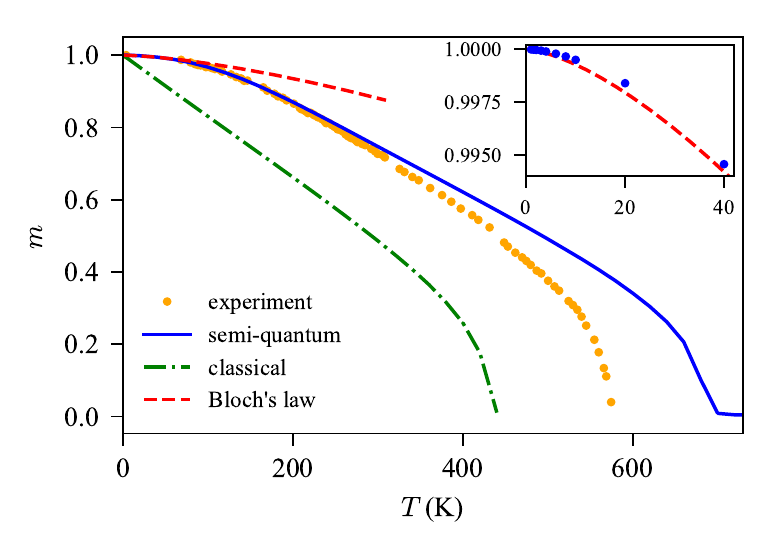}\caption{Temperature
dependent magnetization of YIG calculated using classical and semi-quantum
spin dynamics. The experimental points are from \cite{Anderson1964} and
Bloch's law from Eq.~\eqref{eq:bloch} with $\mathcal{D}=85.2\times{10^{-41}}$
Jm$^{2}$ \cite{Princep2017a}, which in YIG is temperature independent until
close to the Curie temperature. The inset is a close up of the semi-quantum method (blue circles) in the low temperature
regime where Bloch's law (dashed red line) is valid.}
\label{fig:yig_magnetization}%
\end{figure}

Fig.~\ref{fig:yig_magnetization} exposes the obvious problem of classical
statistics to compute magnetizations at low temperatures: The magnetization
decreases much more rapidly with temperature than Bloch's law (and as observed
in experiments). The results with the quantum thermostat, on the other hand,
adhere to Bloch's law for $T<30$~K (see inset) but also agree well with
experiments that signal a breakdown of $T^{3/2}$ scaling, at least until
$\sim300$~K.

The Curie temperatures for the classical ($T_{C}=420$~K) and quantum
thermostated systems ($T_{C}=680$~K) are quite different, while the observed
$T_{C}=550$~K lies between the theoretical values. In contrast to classical
results that obey equipartition, the Curie temperature of quantum approaches
depend on $S$ and we find this also in our semi-quantum approach. For a simple 
ferromagnetic BCC lattice our computed Curie temperatures (not shown) agree well with those
obtained by semi-analytic
approaches~\cite{Oitmaa2004} for a large range of $\mathcal{S}$. The
overestimation of $T_{C}$ compared to the experiment might be caused by exchange parameters that are
slightly too large since the neutron scattering data are fitted only up to
$90$~meV which does not cover the magnon modes with highest energy. Also, the
choice of $\mathcal{S}=5/2$ ($\mu_{s}=5\mu_{B}$) in extracting the exchange
parameters does not fully agree with with measured values of $\mu
_{s,a}=4.11\mu_{B}$ and $\mu_{s,d}=5.37\mu_{B}$ for the octahedral and
tetrahedral sites~\cite{Rodic1999}. Hence, a more accurate set of parameters,
fitted to neutron scattering data for large energy transfers or calculated
from first principles, should solve this discrepancy.  

\paragraph{Heat Capacity}

The magnon heat capacity per unit volume $C_{m}=V^{-1}(\partial U_{m}/\partial
T)_{V}$ is the change in the internal magnetic energy $U_{m}$ with temperature
at constant volume $V$. It can be calculated from the magnon spectrum as
$C_{m}=V^{-1}(\partial/\partial T)\sum_{\vec{k}\nu}\hbar\omega_{\vec{k}\nu
}\langle n_{\vec{k}\nu}\rangle$, where $\langle n_{\vec{k}\nu}\rangle$ is the
Planck distribution. In the low temperature limit magnons occupy only states
close to $k=0$, where the magnon dispersion of ferromagnets is parabolic. For
a single parabolic magnon band \cite{Kittel1963}
\begin{equation}
C_{m}(T)=\frac{1}{V}\frac{5}{8}\frac{{\Gamma\left(  \tfrac{5}{2}\right)
}{\zeta\left(  \tfrac{5}{2}\right)  }}{\pi^{2}}k_{B}\left(  \frac{k_{B}%
T}{\mathcal{D}}\right)  ^{3/2},\label{eq:cm}%
\end{equation}
where $\Gamma(x)$ and $\zeta(x)$ are the Gamma and Riemann zeta functions.

The proportionality $C_{m}\propto T^{3/2}$ should hold for YIG up to energies
of $\hbar\omega_{\vec{k}} / k_B \lessapprox 30~\mathrm{K}$. Rezende and L\'{o}pez Ortiz~\cite{Rezende2015} calculated
the heat capacity for acoustic magnons with finite band-width, but neglected
optical magnons that contribute to the heat capacity at elevated temperatures.
They found that $C_{m}$ saturates at $150$~K, i.e. when the magnon occupation
reaches the upper band edge. 

Here we calculate the heat capacity including all magnon modes and their
interactions. We calculate $C_{m}$ from the energy fluctuations in the
canonical ensemble $\langle U_{m}\rangle_{T}=(1/Z_{m})\sum_{\vec{k}\nu}%
\hbar\omega_{\vec{k}\nu}\mathrm{exp}(-\hbar\omega_{\vec{k}\nu}/k_{B}T),$
where $Z_{m}=\sum_{\vec{k}\nu}\mathrm{exp}(-\hbar\omega_{\vec{k}\nu}/k_{B}T)$
is the partition function. Then $C_{m}=\left(  \langle U_{m}^{2}\rangle
_{T}-\langle U_{m}\rangle_{T}^{2}\right)  /(Vk_{B}T^{2})$, where in a
simulation $\langle\cdots\rangle_{T}$ is an average over a large time interval
at a constant temperature and $V$ is the volume of the system.

\begin{figure}[ptb]
\includegraphics[width=0.48\textwidth]{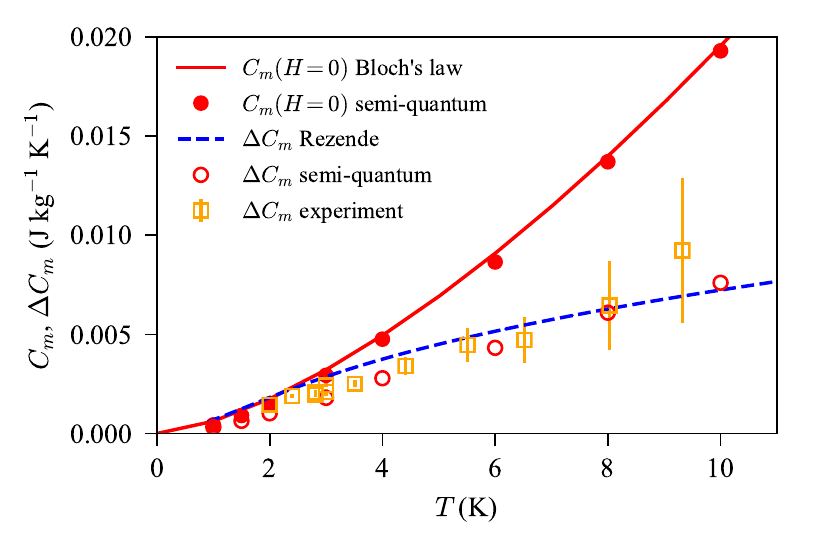}\caption{Low-temperature
magnon heat capacity of YIG calculated with quantum statistics (red circles)
compared to Bloch's law (red solid line).  $\Delta C_{m}=C_{m}(H=0\mathrm{T}%
)-C_{m}(H=7\mathrm{T})$ calculated with quantum statistics (red open
circles)\ is compared with experimental data from Boona and Heremans
Ref.~\onlinecite{Boona2014} (orange open squares) as well as a single magnon
band model.~\cite{Rezende2015} (blue dashed line).}%
\label{fig:yig_heat_capacity_low}%
\end{figure}

Figure~\ref{fig:yig_heat_capacity_low} shows the low temperature region where
the magnon dispersion is, to a good approximation, parabolic and $C_{m}\propto
T^{3/2}$. Calculations using quantum statistics give an excellent agreement
with Bloch's law. The experimental data in
Figure~\ref{fig:yig_heat_capacity_low} have been collected in the range
$T=2-9$~K~\cite{Boona2014}, high enough that dipolar field effects can be
disregarded. The measurements were made by freezing the magnons in a 7 Tesla
field. Even this large field however does not completely remove the magnon
contribution to the heat capacity, especially at the higher end of the
temperature range~\cite{Rezende2015}. To make a proper comparison we repeat
the experimental procedure in our simulation by computing the difference
$\Delta C_{m}=C_{m}(H=0\mathrm{T})-C_{m}(H=7\mathrm{T})$. Our calculations
agree well with the observations as well as the single magnon-band model.

\begin{figure}[ptb]
\includegraphics[width=0.48\textwidth]{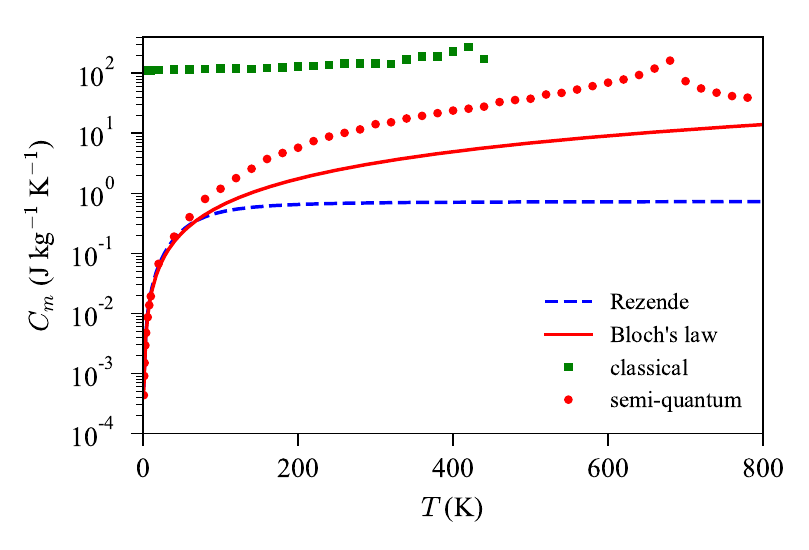}\caption{YIG
magnon heat capacity calculated over a larger temperature range with the
semi-quantum model (red circles), classical model (green squares), compared
with Bloch's law (solid red line) and the single-band model \cite{Rezende2015}
(dashed blue line).}%
\label{fig:yig_heat_capacity}%
\end{figure}

Figure~\ref{fig:yig_heat_capacity} illustrates a pronounced difference between
the classical and semi-quantum models: classical statistics overestimate the
heat capacity by 5 orders of magnitude at low temperatures, and do not depend
on temperature in contrast to the quantum statistical result which approaches
zero like $T^{3/2}$. In spite of this spectacular (and rather obvious)
failure, classical statistics have traditionally been used (and still are) in
both Monte-Carlo and atomistic spin dynamics.

At $T>30$~K non-parabolicities begin and $C_{m}\propto T^{p}$ with power
$p>3/2$. At room temperatures Fig. \ref{fig:yig_heat_capacity} reveals
differences between the approaches of two orders of magnitude. The
finite-width magnon band model \cite{Rezende2015} (dashed line on
figure~\ref{fig:yig_heat_capacity}) saturates prematurely with increasing $T$
because optical and higher acoustic modes become significantly occupied when
approaching room temperature~\cite{Barker2016b}. The parabolic band model
without high-momentum cut-off (Bloch's law) also strongly underestimates
$C_{m}$ because YIG's magnon density of states is strongly enhanced by the
flat bands observed in Fig. 1. The semi-quantum calculation is an order of
magnitude larger than both of these heavily approximated approaches,
benefiting from the complete description of the magnon spectrum as well as
magnon-magnon interactions, while the classical statistics strongly
overestimates the heat capacity up to the Curie temperature. 

\paragraph{Conclusions}

By enforcing Planck statistics for the magnons in the complex ferrimagnet YIG,
we obtain excellent agreement with available inelastic neutron scattering and
magnon heat capacity experiments. Our results prove that fundamental
thermodynamic equilibrium properties can be predicted with confidence when
experimental data are not available, but only when quantum statistics and the
full spin wave spectrum is taken into account. The method is not limited to
YIG or ordered magnets, but can be directly applied to other complex materials 
with local magnetic moments such as spin glasses or paramagnets. 
Our results are a necessary first step to compute non-equilibrium
properties such as magnon conductivities and spin Seebeck coefficients, which
are essential parameters for future applications of magnonic devices.

\section{Acknowledgements}

This work was supported by JSPS KAKENHI Grant No. 26103006, the Graduate
Program in Spintronics (GP-Spin), Tohoku University and DAAD project
`MaHoJeRo'. The authors thank Jiang Xiao, Yaroslav Tserkovnyak, Rembert Duine
and Jerome Jackson for valuable discussion.

\end{document}